\documentstyle[12pt]{article}
\def \a{\alpha}
\def \b{\beta}
\def \l{\lambda}
\def \d{\delta}
\def \p{\partial}
\def \dg{\dagger}
\def \e{\epsilon}

\def \t{\theta}

\def \ra{\rightarrow}
\def \st{\stackrel}
\def \nb{\nabla}
\begin{document}

\centerline{\Large{\bf ON EFFECTS OF GAUGING ON SYMPLECTIC}}
\vskip 0.1in
\centerline{\Large{\bf STRUCTURE, THE HOPF TERM COUPLED TO}}
\vskip 0.1in
\centerline{\Large{\bf $CP^1$ MODEL, AND FRACTIONAL SPIN}}

\vskip 0.5in

\centerline{B.Chakraborty\footnote{e-mail:biswajit@bose.ernet.in} and
A.S.Majumdar\footnote{e-mail:archan@bose.ernet.in}}
\vskip 0.2in

\centerline{S.N.Bose National Centre for Basic Sciences}
\vskip 0.1in
\centerline{Block JD, Sector III, Salt Lake, Calcutta 700091, India}

\vskip 0.5in
\noindent
{\bf Abstract}

We couple the Hopf term to the
relativistic $CP^1$ model and carry out the Hamiltonian analysis at
the classical level. The symplectic structure of the model given
by the set of Dirac Brackets among the phase space
variables is found to be the same as that of the pure $CP^1$ model.
This symplectic structure is shown to be inherited from the global
$SU(2)$ invariant $S^3$ model, and undergoes no modification upon
gauging the $U(1)$ subgroup, except the appearance of an
additional first class constraint generating $U(1)$ gauge transformation.
We then address the question of
fractional spin as imparted by the Hopf term at the classical level.
For that we construct the expression of angular momentum through
both symmetric energy-momentum tensor as well as through Noether's
prescription. Both the expressions agree for the model
indicating no fractional spin is imparted by this term at the classical
level-a
result which is at variance with what has been claimed in the literature.
We provide an argument to explain the discrepancy and
corroborate our argument by considering a radiation gauge fixed
Hopf term coupled to $CP^1$ model, where the desired fractional spin is
reproduced and is given in terms of the soliton number.
Finally, by making the gauge field of the $CP^1$ model dynamical by
adding the Chern-Simons term, the model ceases to be a $CP^1$ model, as
is the case with its nonrelativistic counterpart. This model is also
shown to reveals the existence of `anomalous' spin.
This is however given in terms of the total charge of the system,
rather than any soliton number.
\vskip 0.2in


\pagebreak

\section{Introduction}

Physics of $2+1$ dimensional systems have attracted much attention
in recent years. This is because they provide critical insights
into a large variety of problem in diverse phenomenalogical areas
such as condensed matter physics and quantum gravity~[1,2,3]. In
condensed matter physics one finds applications of these systems,
for example, in quantum Hall effect and anyonic
superconductivity~[2].
The fact that $2+1$ dimensional systems have very
distinctive properties compared to the corresponding higher
dimensional cases is due to the strange nature of the Poincare
group $ISO(2,1)$ in $2+1$ dimensions~[1,3]. For instance, in $2+1$
dimensions there exists the possibility of having fractional spin
and statistics~[3] arising from the occurrence of a multiply connected
configuration space which leads to the presence of nontrivial phase.

These possibilities are realized by coupling the Chern-Simons(CS) or
the Hopf term to various matter fields. For the CS term, apart
from studying the respective Galilean/Poincare covariance, the existence
of fractional spin has been revealed by carrying out a canonical
Hamiltonian analysis in the gauge fixed[4] as well as the gauge
independent[5,6,7] scheme at the classical level, and extending the
analysis to the quantum level. On the other hand, models involving the
Hopf term were initially analysed in the path integral formalism~[3].
For example, Wilczek and Zee(WZ)~[8] had considered the $O(3)$
nonlinear sigma model (NLSM) coupled to the Hopf term. They showed
that the system acquires a nontrivial phase upon an adiabatic rotation
of $2\pi$, signalling existence of fractional spin. Later, the same
system was considered by Bowick et al~[9] where a Hamiltonian analysis
revealed the same fractional spin as obtained by WZ. Although their
analysis was carried out at the quantum level, the fractional spin
they obtained was not a typical quantum effect unlike the WZ case.
This analysis can infact be carried out at the classical level itself
to obtain the same result.
In their gauge
fixed analysis~[9], the existence of fractional spin was demonstrated
by computing the difference between the two expressions of angular
momentum obtained by using the symmetric definition of the energy
momentum(EM) tensor $(J^s)$, and by the Noether prescription $(J^N)$
respectively. In ~[9] $(J^N)$ is just the orbital angular momentum,
as NLSM consists of scalar fields only. One
usually regards the former expression of angular momentum $(J^s)$ as
the physical one, as the corresponding EM tensor is obtained by functional
differentiation of the action with respect to the metric, and is thus
gauge invariant by construction. On the other hand, the latter
expression of angular momentum $J^N$ which usually turns out to be
gauge invariant {\it only} on the constraint surface and that
too under those gauge transformations which reduce to identity
asymptotically~[7].

For the case of the CS term, the above approach has been adopted for
several models~[5-7] to reveal the existence of fractional spin in a
gauge independent manner. The gauge independent scheme has certain
advantages over the corresponding gauge fixed scheme. In the latter
method, the transformation properties of the basic fields under
symmetry transformations get affected by the gauge fixing condition
used. One is thus forced to look at the transformation properties of
{\it gauge invariant} objects to uncover the underlying symmetry. At the
quantum level it therefore becomes rather nontrivial to disentangle the terms
arising due to anomalies from those terms which are artefacts of gauge
fixing conditions. Furthermore the symplectic structure given by  the
set of Dirac Brackets(DB) among the independent phase space variables
usually turn out,
in the gauge fixed scheme, to be more complicated functions of the
fields than their counterparts in the gauge independent scheme.
Quantization by elevating the DB's to quantum commutators in the gauge
fixed scheme (reduced phase space scheme) is therefore liable to possess comparitively more
operator ordering ambiguities than their counterpart in the
gauge independent scheme. The latter scheme corresponds to Dirac
quantization, where the physical states are taken to be gauge
invariant by definition and therefore annihilated by the first
class constraints. These two schemes of quantization are not necessarily
equivalent~[5,10].

It is therefore desirable to have a gauge independent formulation of
the $O(3)$ NLSM coupled to the Hopf term, just as has been done with
models involving the CS term. However, a gauge independent formulation
of models involving the Hopf term is not possible in general. To
understand the difficulty, recall that the conserved current $(\p_{\mu}
j^{\mu} = 0)$ given in terms of the matter fields of any field
theoretical model can be expressed as the curl of a {\it fictitious} gauge
field $a_{\mu}$ $(j^{\mu} \sim \e^{\mu\nu\l}\p_{\nu}a_{\l})$. Coupling
the current $j^{\mu}$ to the field $a_{\mu}$ defines the Hopf term
($\sim j^{\mu}a_{\mu}$). Although this term has formal 
resemblance with the CS term,
for the case of the Hopf term one should be careful not to regard
$a_{\mu}$ as an additional variable in the configuration space.
Rather, $a_{\mu}$ here is determined in terms of $j^{\mu}$, a
procedure which necessitates a gauge fixing condition to be used {\it
a priori} in order to invert the above relation. In this way the Hopf
term is quite distinct to the CS term. Once this inversion is carried
out, the Hopf term becomes a nonlocal expression (quadratic in current
$j^{\mu}$) thereby representing a {\it nonlocal} current-current
interaction. No wonder, in [9] the radiation gauge condition was used
right at the beginning to define the model. It is clear that a gauge
independent analysis of the $O(3)$ NLSM coupled to the Hopf term is
not possible at this stage.

Nevertheless, one can use an alternative $CP^1$ description~[11]
of the same NLSM, which is a $U(1)$ gauge theory having an enlarged
phase space. The advantage here is that unlike in the usual Hopf model
described above, the gauge field, which is the Dirac monopole
connection in the $U(1)$ principle bundle over $S^2$~[12], gets directly
related to the matter fields of the $CP^1$ model in a gauge
independent manner. The current is therefore constructed as a curl of
this gauge field, and finally the Hopf term is obtained by contracting
the current with $a_{\mu}$ as above. Consequently, one can do away
with the inversion and the accompanying gauge fixing condition at any
intermediate step. The Hopf term in the $CP^1$ description thus
becomes a local expression where no gauge fixing condition is required
{\it a priori}.

We thus feel motivated to carry out a gauge independent Hamiltonian
analysis of the $CP^1$ model coupled to the Hopf term, so that Dirac
quantization of the model could eventually be carried out.
In this paper however, we confine our analysis at the classical level,
and look for
the presence of any fractional spin. As mentioned earlier, since the
Hopf term has a formal similarity to the CS term, we would also like
to compare the symplectic structure of the above model with the one
where the gauge filed is given dynamics through the CS term.
While undertaking this job we are
confronted with the following related questions. Since the
$CP^1$ manifold is a
coset space ($CP^1 \sim SU(2)/U(1)$), the $CP^1$ model can be
thought of as obtained by {\it gauging} the $U(1)$ subgroup of
another model enjoying a {\it global} $SU(2)$ symmetry.
So how does
the symplectic structure (Dirac Brackets) of the latter model
compare with that of the $CP^1$ model ? In other words we wish to
investigate the effects of gauging on symplectic structure.
Next, we would also need to know in what way the symplectic
structure gets affected by the presence of the Hopf or CS term.
 
In order  to address the above issues, we begin by presenting
certain mathematical preliminaries towards model building in
section 2. Here we discuss following Balachandran et al,
how the line elements on $S^3$
and $S^2 \sim CP^1$ can be associated with the $SU(2)$ invariant
$S^3$ model and the NLSM or its equivalent
$CP^1$ model respectively. We also discuss how
the Hopf term arises in the $2+1$ dimensional context. We then
carry out the Hamiltonian analysis of the $S^3$ model in section 3.
In section 4 we analyze the symplectic structure of the $CP^1$
model with or without the Hopf term. In the former situation, we
examine the possibility of fractional spin by the explicit
construction of angular momentum through both the Noether
prescription and the symmetric energy momentum tensor, firstly, for
the original $CP^1$ model with the Hopf term, and later for a
gauge fixed version of the Hopf term as well.
In section 5 make the gauge field of
the $CP^1$ Lagrangian dynamical by adding the CS term and study
its effects on the symplectic structure, together with investigating
whether the model retains its $CP^1$ nature or not, in analogy with its
nonrelativistic counterpart~[12]. Section 6 is reserved for some
concluding remarks.

\section{Mathematical preliminaries in model building}

In this section we shall review in detail some of the mathematical
properties of groups and coset spaces to be used to construct a
hierarchy of models whose Hamiltonian analysis shall be carried out in
the subsequent sections. For this, we shall primarily follow [13] where
a general framework for the construction of nonlinear models have been
provided. We will be particularly interested in the group $SU(2)$ and
its coset $CP^1$.
In [12] a method of projection due to
Atiyah~[14]
was used to derive the form of the $U(1)$ connection (monopole
connection) on the $CP^1$ manifold. Here we shall provide an
alternative derivation of the same treating $CP^1$ as a coset space
($CP^1 \sim S^2 \sim SU(2)/U(1)$) where we shall use the techniques
of differential geometry on Lie group manifolds and coset
spaces~[13, 15].
Within this framework we shall also provide a geometrical
interpretation for the equivalence netween the relativistic $CP^1$
model and the $O(3)$ nonlinear sigma model~[11]. We shall also review
the relevant mathematics[15] required for the construction of
the Hopf term.

Let us consider a Lie group $G$ and its subgroup $H$, such that
$G/H$ is a homogeneous coset space. We further assume
that $G/H$ is a symmetric
space implying that the generators $T_{\hat{\alpha}} (\hat{\alpha}
= 1,....,dim[G])$ of $G$ satisfying
$$[T_{\hat{\alpha}}, T_{\hat{\beta}}] = if_{\hat{\alpha}
\hat{\beta}}^{\hat{\gamma}} T_{\hat{\gamma}} \eqno(2.1)$$
can be split into the generators $T_{\bar{\alpha}}$  of $H
(\bar{\alpha} = 1,....,dim[H])$ and the complements $T_{\alpha}
\Bigl(\alpha = 1,....,dim([G] - [H])\Bigr)$ in such a way that
$$[T_{\bar{\alpha}}, T_{\bar{\beta}}] = if_{\bar{\alpha}
\bar{\beta}}^{\bar{\gamma}} T_{\bar{\gamma}}$$
$$[T_{\bar{\alpha}}, T_{\beta}] = if_{\bar{\alpha}
{\beta}}^{\gamma} T_{\gamma}$$
$$[T_{\alpha}, T_{\beta}] = if_{\alpha
\beta}^{\bar{\gamma}} T_{\bar{\gamma}} \eqno(2.2)$$
The rest of the structure constants vanish, i.e.,
$$f_{\bar{\alpha}\bar{\beta}}^{\gamma} =
f_{\bar{\alpha}\beta}^{\bar{\gamma}} = f_{\alpha\beta}^{\gamma} = 0
\eqno(2.3)$$

If $g \in G$, one can construct the following Lie algebra valued
left invariant Maurer-Cartan one form
$$g^{-1}dg = ie^{\hat{\alpha}}T_{\hat{\alpha}} = i(e^{\bar{\alpha}}
T_{\bar{\alpha}} + e^{\alpha}T_{\alpha}) \eqno(2.4)$$
where $e^{\alpha}$ is an orthonormal basis on the cotangent
space over a point in the coset space $G/H$ [15], provided the
generators $T_{\hat{\alpha}}$ are properly normalized~[14].
$e^{\bar{\alpha}}$ the $H$ gauge fields on $G/H$~[13, 17],
and $e^{\hat{\alpha}}$ represents the orthonormal basis on the
group manifold $G$.

We now apply the formalism to the $CP^1$ manifold which is a
symmetric space. The $CP^1$ manifold can also be considered as a
coset space $SU(2)/U(1)$. The Pauli matrices $\sigma_a$'s
($a=1,2,3$) which are the generators of $SU(2)$ satisfy (2.2,2.3).
$\sigma^3$ is the generator of the $U(1)$ subgroup. The $CP^1$
manifold is given by the set of all non-zero complex doublets
$Z = \Biggl(\begin{array}{c} z_1 \\ z_2 \end{array}\Biggr)$
satisfying the normalization condition
$$Z^{\dagger}Z = \vert z_1\vert^2 + \vert z_2\vert^2 =
1\eqno(2.5)$$
and with the identification $Z \sim e^{i\theta}Z$, where
$e^{i\theta} \in U(1)$ is any unimodular number in the complex
plane. (2.5) represents  $S^3$, i.e., the $SU(2)$ group manifold. Thus one
identifies $CP^1$ with the coset space $SU(2)/U(1) \sim S^2$.

Note that given such a normalized doublet $Z =
\Biggl(\begin{array}{c} z_1 \\ z_2 \end{array}\Biggr)$ satisfying
(2.5), one can associate the element
$$g = \pmatrix{z_1 & -z_2^{*} \cr z_2 & z_1^{*}}
\in SU(2)\eqno(2.6)$$
The properly normalized elements for the $SU(2)$ Lie algebra are
the $\sigma_a$'s themselves as they satisfy $tr(\sigma_a\sigma_b)
= 2\delta_{ab}$. Therefore using (2.4) we see that the orthonormal
basis $e_a$'s for the $SU(2)$ group manifold are given by
$$g^{-1}dg = ie^{a}\sigma_a\eqno(2.7)$$
One can write the line element $ds^2$ on the $SU(2)$ group manifold
as
$$ds^2 = e^a e^a\eqno(2.8)$$
which can be simplified  using (2.6) and (2.7) to get
$$ds^2 = - {1\over 2} tr(g^{-1}dg  g^{-1}dg) = {1\over 2}
tr(dg^{-1} dg) = dZ^{\dagger} dZ\eqno(2.9)$$
Following [13], one can associate a nonlinear $S^3$ model to this line
element (2.9),
$${\cal L} = e^a_{\mu}e^{a\mu} \p_{\mu} Z^{\dagger} \p^{\mu} Z\eqno(2.10)$$
enjoying global $SU(2)$ invariance and subject to the constraint (2.5).

The $U(1)$ connection one form $A$ on $CP^1$ is given by $e^3$
which can also be obtained similarly using (2.7) and (2.6) to get
$$A = e^3 = - {i\over 2} tr(g^{-1}dg \sigma_3) = -i Z^{\dagger}dZ\eqno(2.11)$$
This form agrees with the one obtained in [12] by using the method of
projection~[14].

The line element on $CP^1$ is given by
$$d\bar{s}^2 = dM_a dM_a\eqno(2.12)$$
where $M_a$ is a unit $3$-vector, which can be obtained from the
doublet $Z$ by using the Hopf map $M_a = Z^{\dagger} \sigma_a Z$
The $M_a$'s also satisfy
$$g\sigma_3 g^{-1} = M_a \sigma_a\eqno(2.13)$$
where the use of (2.6) has been made.
Using (2.12), (2.7), and (2.13) one can write
$$d\bar{s}^2 = {1\over 2}
tr[d(M_a \sigma_a)d(M_b \sigma_b)] = 4\biggl((e^1)^2 +
(e^2)^2\biggr)\eqno(2.14)$$
Again following [13], one can associate to this line element the model
$${\cal L} = e^a_{\mu}e^{a\mu} \p_{\mu} M_a \p^{\mu} M^a\eqno(2.15)$$
which is precisely the NLSM.

The presence of the factor of $4$ in (2.14) may be understood in
the following manner. Let us parametrize the doublet $Z$ satisfying
(2.5) as
$$Z = \Biggl(\begin{array}{c} z_1 \\ z_2 \end{array}\Biggr) =
\Biggl(\begin{array}{c} e^{i\alpha} Cos\phi \\ -e^{-i\beta} Sin\phi
\end{array}\Biggr)\eqno(2.16a)$$
so that the associated $SU(2)$ group element (2.6) is
$$g = \pmatrix{e^{i\alpha}Cos\phi &
e^{i\beta}Sin\phi \cr -e^{-i\beta}Sin\phi & e^{-i\alpha}Cos\phi}
\in SU(2)\eqno(2.16b)$$
Substituting in (2.7) one can solve for $e^a$'s to get
$$e^1 = Sin\phi Cos\phi Cos(\alpha - \beta)d\alpha +
Sin\phi Cos\phi Cos(\alpha - \beta)d\beta -Sin(\alpha -
\beta)d\phi$$
$$e^2 = Sin\phi Cos\phi Sin(\alpha - \beta)d\alpha +
Sin\phi Cos\phi Sin(\alpha - \beta)d\beta + Cos(\alpha -
\beta)d\phi$$
$$e^3 = Cos^2\phi d\alpha + Sin^2\phi d\beta\eqno(2.17)$$

The $e^a$'s and the $M_a$'s, parametrized by the
polar coordinates $\Theta$ and $\Phi$ [12], and constructed through the
Hopf map, are given in the gauge $z_2^{*} = z_2 (\beta =0)$ by
$$e^1 = Sin\phi Cos\phi Cos\alpha d\alpha - Sin\alpha
d\phi $$
$$e^2 = Sin\phi Cos\phi Sin\alpha d\alpha + Cos\alpha d\phi$$
$$e^3 = Cos^2\phi d\alpha\eqno(2.18a,b,c)$$
$$M_1 = -Sin2\phi Cos\alpha = Sin\Theta Cos\Phi$$
$$M_2 = Sin2\phi Sin\alpha = Sin\Theta Sin\Phi$$
$$M_3 = Cos2\phi = Cos\Theta\eqno(2.19a,b,c)$$
From (2.19) the variables $\Theta$ and $\Phi$ can be easily
identified as
$$\Theta = -2\phi ; \Phi = -\alpha\eqno(2.20)$$

Substituting these back in (2.18c), one gets the connection
one-form, valid for the `southern' hemisphere as
$$A^{(-)} = e^3 = {1\over 2}(-1-Cos\Theta)d\Phi\eqno(2.21a)$$
Proceeding similarly for the gauge $z_1 = z_1^{*} (\alpha = 0)$,
one gets the corresponding expression for the connection one-form ,
valid for the `northern' hemisphere, as
$$A^{(+)} = e^3 = {1\over 2}(1-Cos\Theta)d\Phi\eqno(2.21b)$$
$A^{(+)}$ and $A^{(-)}$, related by the gauge transformation
$(A^{(+)} - A^{(-)} = d\Phi)$, define the Dirac magnetic
monopole configuration
on $CP^1 \sim S^2$.

Now it is easy to see that in the gauge $z_2^{*} = z_2$, $e^1$ and
$e^2$ satisfy
$$(e^1)^2 + (e^2)^2 = {1\over 4}[Sin^2\Theta d\Phi^2 +
d\Theta^2]\eqno(2.22)$$
where the use of (2.20) has been made. (2.22) has an overall factor
of `$1/4$', and when substituted in (2.14) indeed produces the line
element of $S^2$. A similar result holds for the other gauge
$z_1^{*} = z_1$ also. The lesson is that although the $e^a$'s in
(2.8) provide an orthonormal basis on $S^3$, the relevant $e^a$'s
restricted to $S^2$, i.e., $e^1\vert_{\beta = 0}$ and
$e^2\vert_{\beta = 0}$ are not orthonormal. Rather, $\biggl(
2e^1\vert_{\beta = 0}, 2e^2\vert_{\beta = 0}\biggr)$ provide an
orthonormal basis on $S^2$. The factor of `$2$' can be traced to the
first relation involving $\Theta$ and $\phi$ in (2.20). The line
element (2.12) on $S^2$, written in terms of the gauge invariant
$M_a$ variables, automatically gets $e^1$ and $e^2$
projected onto the cotangent space of $S^2$. Finally, we note that
the line element on $S^2$ (2.14) can be expressed alternatively
using (2.8) as
$$d\bar{s}^2 = 4[ds^2 - (e^3)^2]\eqno(2.23)$$
Further, using (2.5) and (2.11), one can rewrite (2.23) as
$${1\over 4}dM_a dM_a = [dZ^{\dagger}dZ -(A)^2] = (dZ -
iAZ)^{\dagger}(dZ - iAZ)\eqno(2.24)$$

Now, this implies that we have the following identity
$${1\over 4}\partial_{\mu}M_a \partial^{\mu}M_a =
(D_{\mu}Z)^{\dagger} (D^{\mu}Z)\eqno(2.25a)$$
where $D_{\mu} = (\partial_{\mu} - iA_{\mu})$ stands for the
covariant derivative operator, and
$$A_{\mu} = -i Z^{\dagger} \partial_{\mu}Z\eqno(2.25b)$$
is the $U(1)$ gauge field obtained by pulling back the connection
(2.11) on ${\cal M}$.
The right hand side of (2.25a) corresponds to the $CP^1$ model having a
local $U(1)$ invariance. From (2.25) it is clear
that Lagrangian densities of the $CP^1$ model and the
NLSM (2.15) are the same. Note that both the NLSM and the $CP^1$
Lagrangian (upto a factor) have been obtained from the same line element
(2.12) or its equivalent (2.23) using the prescription of [13].
Thus these two models are classically
equivalent~[11]. Note that we do not have any dynamical term for the
gauge field in the $CP^1$ model.

We have thus constructed physical models associated with the line
elements of $S^3$ and its coset $S^2$.
Let us now address the issue of the existence of solitonic
configurations~[13, 16] for the models constructed above.  For the case of
the $O(3)$ NLSM, it is necessary for the fields
$M_a$ to tend to a constant configuration asymptotically for finite
energy static solutions to exist. With this requirement, the
two-dimensional plane ${\cal D}$  gets effectively
compactified to $S^2$, so that the configuration space ${\cal C}$ is the
set of all maps $\{f\}$:
$$f : S^2 \rightarrow S^2 (field \> manifold)\eqno(2.26)$$
Clearly, the splits into a disjoint union of path connected spaces~[16] as
$$\Pi_0(C) = \Pi_2(S^2) = Z\eqno(2.27)$$
Hence, there exist solitons or skyrmions in this model,
characterized by the set of integers ${\cal N} \in Z$ given by
$${\cal N} = \int d^2x j^0 (x)\eqno(2.28)$$
where $j^0$ is the time component of the identically conserved
current $(\partial_{\mu}j^{\mu} = 0)$ given by
$$j^{\mu} = {1\over 8\pi} \epsilon^{\mu\nu\lambda} \epsilon_{abc}
M_a\partial_{\nu}M_b \partial_{\lambda} M_c\eqno(2.29)$$
Note that the conservation of $j^{\mu}$ holds irrespective of any
equation of motion. ${\cal N}$, referred to as the soliton number, labels
disconnected pieces of the configuration space $C$.
Parametrizing the unit vector $M_a$ by polar coordinates ($\Theta ,
\Phi$) as in (2.19), $j^{\mu}$ can be expressed as the curl of the
gauge fields $A_{\mu}$ as
$$j^{\mu} = {1\over 2\pi} \epsilon^{\mu\nu\lambda} \partial_{\nu}
A_{\lambda}\eqno(2.30)$$
where the use of (2.21) has been made.

In any soliton number sector, the fundamental group of the
configuration space $C$ is nontrivial~[16] since
$$\Pi_1(C) = \Pi_3(S^2) = Z\eqno(2.31)$$
This implies that loops based at any point in the
configuration space fall into separate homotopy classes labelled by
another integer $H$. This integer can be given a representation by
the so called Hopf term~[3]
$$H = \int j^{\mu} A_{\mu} d^3x\eqno(2.32)$$
This term can be added to the NLSM
to impart fractional spin and statistics to the
solitons~[8,9] - a possibility arising out of (2.31). Note that in adding
the Hopf term (2.32), we do not enlarge the configuration space.
This is because the $A_{\mu}$ field is not treated as an
independent variable in the configuration space; rather is
determined in terms of $j_{\mu}$ by inverting (2.30). This is the
reason why the Hopf term provides a nonlocal current-current
interaction in general. However, in the $CP^1$ version the Hopf term
is local since $A_{\mu}$ (2.36b), and hence $j_{\mu}$ (2.30)
are given in terms of
local expressions of the $Z$ fields. Making use of (2.25b) and
(2.30) in (2.32), one gets the following expression for the Hopf term in the
hermitian form
$$H = -{1\over 4\pi} \int d^3x \epsilon^{\mu\nu\lambda} Z^{\dagger}
\stackrel{\leftrightarrow}{\partial_{\mu}}Z \partial_{\nu}
Z^{\dagger} \partial_{\lambda} Z\eqno(2.33)$$
This Hopf term can be expressed alternatively in terms of the
Cartan-Maurer form (2.4) $\biggl(\sim \int_M tr(g^{-1}dg)^3\biggr)$ as
has been shown in [13].

In the following two sections we shall consider the $S^3$ model and
the $CP^1$ model extended by the Hopf term respectively, and carry
out their Hamiltonian analysis. Apart from studying the
similarities in their symplectic structure, our aim is to look for
any fractional spin generated by the Hopf term.

\section{Hamiltonian analysis of the $S^3$ model}

The $S^3$ model is given by the Lagrange density
$${\cal L} = \p_{\mu}Z^{\dg}\p^{\mu}Z - \l(Z^{\dg}Z -1)\eqno(3.1)$$
This is obtained from (2.10) by incorporating the constraint
$(Z^{\dg}Z = 1)$ (2.5) by the Lagrange multiplier $\l$. The model
posseses a global $SU(2)$ invariance but no local symmetry. The
Legendre transformed Hamiltonian ${\cal H}_c$ is given by
$${\cal H}_c = \pi^{*}_{\a}\pi_{\a} + \p_iZ^{\dg}\p_iZ +
\l(Z^{\dg}Z -1)\eqno(3.2)$$
The canonically congugate momenta of the configuration space
variables $z_{\a}$, $z^{*}_{\a}$ and $\l$are given by
$$\pi_{\a} = {\d {\cal L} \over \d \dot{z}_{\a}} =
\dot{z}^{*}_{\a} \eqno(3.3a)$$
$$ \pi^{*}_{\a} = {\d {\cal L} \over \d \dot{z}^{*}_{\a}} =
\dot{z}_{\a} \eqno(3.3b)$$
$$\pi_{\l} = {\d {\cal L} \over \d \dot{\l}} = 0\eqno(3.3c)$$
respectively. Thus the only primary constraint of the model given
by
$$\pi_{\l} \approx 0\eqno(3.4)$$
Time preserving (3.4) with respect to the Hamiltonian (3.2) we get
the secondary constraint
$$C_1(x) = Z^{\dg}Z -1 \approx 0\eqno(3.5)$$
Repeating the above procedure, we obtain the following tertiary
constraint
$$C_2(x) = \pi^{*}_{\a}z^{*}_{\a} + \pi_{\a}z_{\a} \approx
0\eqno(3.6)$$

It is apparent that the constraints (3.5) and (3.6) are second
class as the Poisson Bracket (PB)
$$C_{ij}(x,y) = \{C_i(x), C_j(y)\} = 2\e_{ij}\d (x-y)\eqno(3.7a)$$
does not vanish. The inverse of this matrix is given by
$$(C^{-1})_{ij} (x,y) = -{1\over 2} \e_{ij} \d (x-y)\eqno(3.7b)$$
Clearly, there exist no more constraints. The
second class constraints can now be `strongly' implemented by using
the appropriate Dirac Brackets (DB). The DB between two quantities
$A(x)$ and $B(y)$ is given by~[10]
$$\{A(x), B(y)\} = \{A(x), B(y)\}_{PB} - \int du dv \{A(x),
C_i(u)\}(C^{-1})_{ij} (u,v) \{C_j(v), B(y)\}\eqno(3.8)$$
Using (3.7b) and (3.8) one finds the following DB's among the phase
space variables
$$\{z_{\a}(x), \pi_{\b}(y)\} = (\d_{\a\b} - {1\over 2}
z^{*}_{\b} z_{\a}) \d (x-y)$$
$$\{\pi_{\a}(x), \pi_{\b}(y)\} =  {1\over 2}
(\pi_{\a}z^{*}_{\b} - z_{\a}^{*}\pi_{\b}) \d (x-y)$$
$$\{z_{\a}(x), z_{\b}(y)\} = \{z_{\a}^{*}(x), z_{\b}(y)\} = 0$$
$$\{z_{\a}(x), \pi^{*}_{\b}(y)\} = -{1\over 2} z_{\a}(x) z_{\b}(x) \d
(x-y)$$
$$\{\pi_{\a}(x), \pi^{*}_{\b}(y)\} = {1\over 2} (\pi_{\a} z_{\b} -
\pi^{*}_{\b} z^{*}_{\a}) \d (x-y)\eqno(3.9)$$
This provides us with the symplectic structure of this model.

With these DB's the total Hamiltonian can now be written as
$${\cal H} = \pi^{\dg}\pi + \p_i Z^{\dg} \p_i Z + u \pi_{\l}
\eqno(3.10)$$
where $u$ is an arbitrary Lagrange multiplier for the constraint
(3.4). This constraint is the only first class constraint of the
model (3.1). It gives vanishing brackets with all the phase space
variables of the model except the Lagrange multiplier $\l$. The
time evolution of $\l$ is therefore given by
$$\dot{\l} = \int d^2x \{\l , {\cal H}\} = u\eqno(3.11)$$
which is again arbitrary.  Without loss of generality, one can
therefore put
$$\l = 0\eqno(3.12)$$
Clearly, (3.4) along with (3.12) form a new second class pair which
is strongly implemented by an additional DB
$$\{\l, \pi_{\l}\} = 0\eqno(3.13)$$
So finally the Hamiltonian reduces to
$${\cal H} = \pi^{\dg} \pi + \p_i Z^{\dg} \p_i Z\eqno(3.14)$$
Note that the model has no non-trivial first class constraint, and
hence, no gauge symmetry. This is expected, since the Lagrangian
(3.1) has only a global $SU(2)$ symmetry.

\vskip 0.5in  

\section{The $CP^1$ model with or without the Hopf term}

In this section we perform a Hamiltonian analysis of the
$CP^1$ model coupled to the Hopf term, the essential features of
which have been outlined in [18]. In addition to working out the
complete constraint algebra, here we also wish to comment on some
technical as well as physical subtleties in the calculation.
We also present the corresponding analysis for the pure $CP^1$
case. Our aim is to highlight any difference in the symplectic
structure made by the presence of the Hopf term, and also to
investigate the existence of fractional spin in the model. Towards
the end of this section we consider the example of a radiation
gauge-fixed
Hopf-$CP^1$ model where fractional spin is revealed by the
construction of the angular momentum operator.

\subsection{Constraint analysis of the model}

The model is given by the Lagrangian
$${\cal L} = {\cal L}_0 + {\cal L}_H - {\l}(Z^{\dg}Z -1)
\eqno(4.1a)$$
where
$${\cal L}_0 = (D_{\mu}Z)^{\dg}(D^{\mu}Z)\eqno(4.1b)$$
corresponds to the right hand side of (2.25a) and
$${\cal L}_H = \theta \epsilon^{\mu\nu\lambda} \left[Z^{\dagger}
\partial_{\mu}Z\partial_{\nu}Z^{\dagger}\partial_{\lambda}Z +
\partial_{\mu}Z^{\dagger}Z\partial_{\lambda}Z^{\dagger}\partial_{\nu}Z
\right]\eqno(4.1c)$$
is the Hopf term corresponding to (2.33) with strength $\theta$.
This Hopf term was coupled  to the nonrelativistic version of the $CP^1$
model in [19] where it was shown to alter the spin algebra.
The configuration space variables are $z_{\a}, z^{*}_{\a}, A_i,
A_0,$ and ${\l}$. The corresponding momenta are
$$\pi_{\alpha}  =  {\delta {\cal L}\over \delta \dot{z}_{\alpha}} =
(D_0z_{\a})^{*} + {\d {\cal L}_H \over \d \dot{z}_{\a}} =
(D_0z_{\alpha})^{*} + \theta \epsilon^{ij} \left[\partial_i
Z^{\dagger}\partial_jZz^{*}_{\alpha} + Z^{\dagger}\partial_iZ
\partial_jz^{*}_{\alpha} - \partial_iZ^{\dagger}Z\partial_j
z^{*}_{\alpha}\right] \eqno(4.2)$$
$$\pi^{*}_{\alpha}  =  {\delta {\cal L}\over \delta \dot{z}^{*}_{\alpha}} =
(D_0z_{\a}) + {\d {\cal L}_H \over \d \dot{z}^{*}_{\a}} =
(D_0z_{\alpha}) + \theta \epsilon^{ij} \left[-Z^{\dagger}\partial_i
Z\partial_jz_{\alpha} + \partial_jZ^{\dagger}
\partial_iZz_{\alpha} + \partial_iZ^{\dagger}Z\partial_j
z_{\alpha}\right] \eqno(4.3)$$
$$\pi_i  =  {\delta {\cal L}\over \delta \dot{A}^i} = 0
\eqno(4.4)$$
$$\pi_0  =  {\delta {\cal L}\over \delta \dot{A}^0} = 0
\eqno(4.5)$$
$$\pi_{\lambda}  =  {\delta {\cal L}\over \delta \dot{\lambda}} =
0\eqno(4.6)$$

The Eqs.(4.4-4.6) represent the primary constraints of this model.
Note that ${\cal L}_H$ contains terms   which are first order in the time
derivative involving either $\dot{z}_{\alpha}$ or
$\dot{z}^{*}_{\alpha}$. One can thus write
$${\d {\cal L}_H \over \d \dot{z}_{\a}} \dot{z}_{\a} +
{\d {\cal L}_H \over \d \dot{z}^{*}_{\a}} \dot{z}^{*}_{\a} = {\cal
L}_H \eqno(4.7)$$
Using this the Legendre transformed Hamiltonian can be obtained as
$${\cal H}_c = \pi^{*}_{\a} \dot{z}^{*}_{\a} + \pi_{\a}
\dot{z}_{\a} - {\cal L}$$
$$ = (D_0Z)^{\dg}\dot{Z} + \dot{Z}^{\dg} (D_0Z) - (D_{\mu} Z)^{\dg}
(D^{\mu}Z) + {\l}(Z^{\dg}Z - 1)\eqno(4.8)$$

Writing (4.8)  in terms of the phase space variables, one gets
$${\cal H}_c = \pi_{\alpha}^{*}\pi_{\alpha} - iA_0\biggl(\pi_{\alpha}^{*}
z_{\alpha}^{*} - \pi_{\alpha}z_{\alpha} + z_{\alpha} {\delta {\cal
L}_H \over \delta\dot{z}_{\alpha}} - z_{\alpha}^{*} {\delta {\cal
L}_H \over \delta\dot{z}_{\alpha}^{*}}\biggr)$$
$$ - \biggl(\pi_{\alpha} {\delta {\cal
L}_H \over \delta\dot{z}_{\alpha}^{*}} + \pi_{\alpha}^{*}{\delta {\cal
L}_H \over \delta\dot{z}_{\alpha}}\biggr) + {\delta {\cal
L}_H \over \delta\dot{z}_{\alpha}}{\delta {\cal
L}_H \over \delta\dot{z}_{\alpha}^{*}} + \vert D_i Z\vert^2 + \lambda\bigl(
Z^{\dagger}Z - 1\bigr)\eqno(4.9)$$

Preservation of the primary constraints (4.4-4.6) in time yield the following
set of secondary constraints
$$A_i + {i\over 2(Z^{\dagger}Z)}
Z^{\dagger}\stackrel{\leftrightarrow}{\partial_i} Z \approx 0
\eqno(4.10)$$
$$i\biggl(\pi^{*}_{\alpha}z^{*}_{\alpha} - \pi_{\alpha}z_{\alpha}
+ z_{\alpha}{\delta
{\cal L}_H \over \delta\dot{z}_{\alpha}} - z^{*}_{\alpha}{\delta {\cal
L}_H \over \delta\dot{z}_{\alpha}^{*}}\biggr) \approx 0 \eqno(4.11)$$
$$Z^{\dagger}Z - 1 \approx 0\eqno(4.12)$$
respectively. From the constraint (4.12), a new tertiary
constraint
$$\pi^{*}_{\alpha}z^{*}_{\alpha} + \pi_{\alpha}z_{\alpha} \approx
0\eqno(4.13)$$
is obtained. The constraint (4.11) can be simplified further using
(4.13) to yield
$$i(\pi^{*}_{\alpha}z^{*}_{\alpha} - \pi_{\alpha}z_{\alpha} +
2\theta\epsilon^{ij}\partial_iZ^{\dagger}\partial_jZ) \approx
0\eqno(4.14)$$
Finally, by demanding the preservation of (4.14) in time,
one more constraint
$$\pi^{*}_{\alpha}\pi_{\alpha} + \left(D_iD_iZ\right)^{\dagger}Z - \lambda +
{\theta}-dependent \>\> terms \approx 0\eqno(4.15)$$
is obtained,
where the last $\theta$-dependent terms are independent of
$\lambda$. It can be checked that there exist no further constraints.

At this stage it is necessarry to classify the total set of
constraints (4.4-4.6, 4.10, 4.12-4.15) into
first class and second class contraints. A comparison with the
$S^3$ model discussed in the previous section shows that the
constraints (4.6, 4.12 and 4.13) are similar to the constraints
(3.4, 3.5 and 3.6) respectively. In addition here we have the
constraints (4.4, 4.5 and 4.14). The constraint (4.15) corresponds
to the condition ${\l} = 0$ (3.12) of the previous section. The
appearence of (4.4 and 4.5) is in keeping with the fact that
we have introduced a
background gauge field $A_{\mu}$ to gauge the $U(1)$ subgroup of
the global $SU(2)$ group of the $S^3$ model. The other nontrivial
constraint (4.14) is therefore expected to be the Gauss constraint.
Besides, (4.14) is obtained by preserving the constraint (4.5) in
time, just as in Maxwell electrodynamics~[10].
It can be easily seen that (4.4, 4.10),
(4.6,4.15), and (4.12,4.13) form pairs of second class constraints.
Only the constraint (4.14) is first class, leaving apart the trivial
constraint (4.5).
The first two pairs are  `strongly'
implemented by the (DB)
$$\{A_i(x), \pi^j(y)\} = 0 \eqno(4.16)$$
$$\{\lambda(x), \pi_{\lambda}(y)\} = 0\eqno(4.17)$$
These are the additional DB's that we get in this model other
than those in (3.9) obtained from the last pair (4.12, 4.13).

The constraint  (4.14) when simplified further using (4.13) gives
$$G(x) \equiv i\biggl(2\pi_{\alpha}(x)z_{\alpha}(x) -
2\theta\epsilon^{ij}\partial_iZ^{\dagger}(x)\partial_jZ(x)\biggr) \approx
0\eqno(4.18)$$
which can be shown using (3.9) to generate a $U(1)$ gauge transformation
$$\delta z_{\alpha}(x) = \int d^2y f(y) \{z_{\alpha}(x), G(y)\}
= if(x)z_{\alpha}(x)\eqno(4.19)$$
Therefore (4.18) can be identified with the Gauss constraint. The
transformation property of the momenta variables $\pi_{\a}$ can be
obtained either through  the DB (3.9), or else by using the basic
transformation properties of the fundamental fields $z_{\a}$ (4.19)
to get
$$\d \pi_{\a} (x) = \int d^2y f(y) \{\pi_{\a}(x), G(y)\} $$
$$ = i[f(x) \pi_{\a}(x) - 2\theta \e_{ij} \p_i f(D_j z_{\a}^{*})]
\eqno(4.20)$$
Note that the Gauss constraint (4.18) was absent in the $S^3$
model where the ${\l} = 0$ condition was put in by hand. In
contrast, here the constraint (4.15) can be  obtained by preserving the
Gauss constraint in time. The exact form of ${\l}$ is
inconsequential for the DB (4.17) and the Hamiltonian since ${\l}$
being a Lagrangian multiplier, enforces the `strongly' valid second
class constraint (4.12).

The final form of the total Hamiltonian is given by
$${\cal H} = \pi^{*}_{\a} \pi_{\a} - \biggl(\pi_{\a} {\d {\cal L}_H
\over \d \dot{z}^{*}_{\a}} + \pi^{*}_{\a} {\d {\cal L}_H \over \d
\dot{z}_{\a}}\biggr) + {\d {\cal L}_H \over \d \dot{z}^{*}_{\a}}
{\d {\cal L}_H \over \d \dot{z}_{\a}} + \vert D_iZ \vert^2 - A_0G
\eqno(4.21)$$
Note that since (4.10) and (4.12) are second class constraints, and
therefore are `strongly' valid, one can simplify (4.10) to write
$$A_i = -i Z^{\dg}\p_i Z\eqno(4.22)$$
$A_i$ therefore ceases to be an independent degree of freedom.
Note also, that the Gauss constraint (4.18) can be expressed
alternatively using (4.2) and (4.3) as
$$Z^{\dg}(D_0 Z) - (D_0 Z)^{\dg} Z \approx 0\eqno(4.23)$$
which can be solved for $A_0$ to yield
$$A_0 = -i Z^{\dg} \p_0 Z\eqno(4.24)$$
However, (4.22) is not a constraint equation as it involves a time
derivative, unlike (4.24). Nevertheless, it could sometimes be useful to
write (4.22 and 4.24) compactly as
$$A_{\mu} = -i Z^{\dg} \p_{\mu} Z\eqno(4.25)$$
which is just the expression (2.25b) obtained by the geometrical
approach discussed in section 2. The consistency of the whole
approach is apparent since the $CP^1 \simeq S^2$ manifold admits a
unique $U(1)$ connection up to a gauge transformation. This follows
from the fact that the first deRham cohomology group vanishes for
the $CP^1$ manifold $\Bigl(H^1(CP^1) = 0\Bigr)$~[15].

After having studied the $CP^1$ model coupled to the Hopf term, let
us now consider the case when the Hopf term does not exist, i.e.,
if $\theta = 0$ in (4.1a). The Lagrangian for the pure $CP^1$ model
is given by
$${\cal L} = (D_{\mu} Z)^{\dg} (D^{\mu} Z) - {\l} (Z^{\dg}Z -
1)\eqno(4.26)$$
In this case the symplectic structure can easily be shown to remail
essentially the same as in the case with the presence of the Hopf
term. However, the canonically conjugate momenta variables (i.e.,
the counterparts of (4.2 and 4.3)) are different, and are given by
$$\pi_{\a} = (D_0Z)^{*}_{\a}\eqno(4.27)$$
$$\pi^{*}_{\a} = (D_0Z)_{\a}\eqno(4.28)$$
The other momenta variables remain the same as (4.4 - 4.6). The
secondary and the tertiary constraints following from the Legendre
transformed Hamiltonian
$${\cal H}_c = \pi^{*}_{\a} \pi_{\a} - i A_0 (\pi^{*}_{\a}
z^{*}_{\a} - \pi_{\a}z_{\a}) + \vert D_iZ \vert^2 + {\l} (Z^{\dg}Z
-1)\eqno(4.29)$$
which do not undergo any change are given by (4.10, 4.12 and 4.13).
However, the Gauss constraint (i.e., the counterpart of (4.14 or
4.18) now becomes
$$G(x) \equiv i\biggl(\pi^{*}_{\a}(x)z^{*}_{\a}(x) - \pi_{\a}(x)
z_{\a}(x)\biggr) = 2i\pi_{\a}(x)z_{\a}(x) \approx 0\eqno(4.30)$$
The preservation of the Gauss constraint in time yields
$$\pi^{*}_{\a} \pi_{\a} + (D_iD_iZ)^{\dg}Z - {\l} \approx
0\eqno(4.31)$$
The final symplectic structure is given by the DB's (3.9, 4.16 and
4.17) and therefore undergoes no change. The total Hamiltonian
obtained from (4.29, and 4.30) is given by
$${\cal H} = \pi^{*}_{\a} \pi_{\a} + \vert D_iZ \vert^2 -
A_0G\eqno(4.32)$$
Finally, the expression for the gauge field $A_{\mu}$ (4.25) in
terms of the matter fields $Z$ holds together with the `strongly' valid
relation (4.22) for the spatial components.
Thus it is clearly evident
that the presence or absence of the Hopf term has no effect on the
symplectic structure.

\subsection{Angular momentum in the $CP^1$ model}

Here we shall construct various spacetime symmetry generators of
the model (4.1) obtained from both the Noether's prescription, and
the symmetric expression of the energy momentum (EM) tensor. The
latter can be obtained by functionally differentiating the action
with respect to the spacetime metric. We shall focus particularly
on the angular momentum since our goal is to look for any
fractional spin generated by the presence of the Hopf term. At this
stage it needs to be mentioned that fractional spin
was found to be induced by the Hopf term in the equivalent
nonlinear sigma model (NLSM) by Bowick et al~[9]. However, it is
important to note that in [9] a gauge had to be fixed right at the
beginning in order to express the gauge field $A_{\mu}$ in terms of
the current $j_{\mu}$ (see (2.43)). On the other hand, in the
$CP^1$ version, no gauge fixing is necessary, and one can perform a
gauge independent Hamiltonian analysis.

The symmetric expression for the energy momentum (EM) tensor is given by
$$T^s_{\mu\nu} = (D_{\mu}Z)^{\dagger}(D_{\nu}Z) +
(D_{\nu}Z)^{\dagger}(D_{\mu}Z) - g_{\mu\nu}(D_{\rho}Z)^{\dagger}
(D^{\rho}Z)\eqno(4.33)$$
Note that the $\theta$- dependent term ${\cal L}_H$ does not
contribute to this expression since it is a topological term which
is independent of the metric. It follows that the symmetric
expression for the Hamiltonian is given by
$$H^s = \int d^2x [2(D_0Z)^{\dg}(D_0Z) -
(D_{\mu}Z)^{\dg}(D^{\mu}Z)]\eqno(4.34)$$
which can be rewritten using (4.8 and 4.12) as
$$H^s = \int d^2x \biggl({\cal H}_c - i A_0 [(D_0Z)^{\dg}Z -
Z^{\dg}(D_0Z)]\biggr)\eqno(4.35)$$
It can be checked that (4.35) generates the appropriate time
translation. The last expression reduces to
$$H^s \approx \int d^2x {\cal H}_c\eqno(4.36)$$
on the Gauss constraint surface.

Let us  now consider the momentum generator obtained from
(4.33) given by
$$P^s_i  = \int d^2x T^s_{0i} = \int d^2x
[(D_0Z)^{\dg}(D_iZ) + (D_iZ)^{\dg}(D_0Z)]\eqno(4.37)$$
This expression can be simplified further to get
$$P^s_i = P^N_i + 2i\theta \e^{jk} \int d^2x [A_j \p_i Z^{\dg} \p_k
Z - A_i \p_j Z^{\dg} \p_k Z - A_j \p_k Z^{\dg} \p_i Z] - \int
d^2xA_i(x)G(x)\eqno(4.38)$$
where
$$P^N_i = \int d^2x p^N_{0i} =
\int d^2x [\pi_{\a} \p_i z_{\a} + \pi^{*}_{\a} \p_i
z^{*}_{\a}]\eqno(4.39)$$
is the expression of momentum obtained from the Noether theorem.
The presence of the second $\t$ -dependent term is a reflection of the
fact that the canonically conjugate momentum variables $\pi_{\a}$ (4.2)
and $\pi^*_{\a}$ (4.3) get a $\t$ -dependent contribution from the Hopf
term, over the corresponding variables (4.27) and (4.28) in the pure
$CP^1$ case.
Now using the fact that in two spatial dimensions one can write
$\p_i A_j - \p_j A_i = \e_{ij} B$ ($B$ being the magnetic field), it
can be shown that the $\theta$---dependent term in (4.38) vanishes.
However because of the presence of the last term involving the Gauss
constraint $G(x)$ in (4.38), $P_i^s$ fails to generate appropriate
translations, i.e.,
$$\{z_{\a}(x), P_i^s\} = D_iz_{\a}\eqno(4.40)$$
in contrast to $P^N$ (4.39) which by construction generates the
appropriate translation
$$\{z_{\a}(x), P_i^N\} = \p_iz_{\a}\eqno(4.41)$$

However, note that one has the liberty to modify the EM tensor (4.33)
by an appropriate linear combination of first class constraint(s)
(here only (4.18)) with tensor valued coefficients $u_{\mu\nu}$
$${\tilde T}^s_{\mu\nu} = T^s_{\mu\nu} +
u_{\mu\nu}G\eqno(4.42)$$
By looking at the expression (4.38) it is clear that with the choice
$$u_{0i} = A_i\eqno(4.43)$$
one gets
$${\tilde T}^s_{0i} = T^N_{0i}\eqno(4.44)$$
and correspondingly
$${\tilde P}_i^s \equiv \int d^2x {\tilde T}^s_{0i} =
P_i^N\eqno(4.45)$$
which generates the appropriate translation $\{z_{\a},
{\tilde P}_i^s\} = \p_iz_{\a}$ just as (4.41), and can therefore
be identified as momentum. Note that this way of obtaining the
modified expression ${\tilde T}^s_{0i}$ from $T^s_{0i}$ is tantamount
to simplifying $T^s_{0i}$ on the Gauss constraint surface.

Nextly, one can write down the two expressions of angular momentum
as
$$J^s = \int d^2x \epsilon_{mj} x_m {\tilde T}^s_{0j}$$
$$J^N = \int d^2x \epsilon_{mj} x_m T^N_{0j}\eqno(4.46)$$
Note that the Noether expression $J^N$ corresponds only to the
orbital angular momentum since we are dealing with scalar fields.
Apart from the fact that they generate appropriate rotation, one
can easily see in view of (4.44) that these two expressions match
exactly. Thus
$$J^s = J^N\eqno(4.47)$$
Usually, $J^s$ is regarded as the physical angular momentum as it
is obtained from the symmetric expression of the EM tensor which is
gauge invariant by construction. On the other hand $J^N$ is usually
found to be gauge invariant only on the Gauss constraint surface,
and that too only under those gauge transformations that reduce to
identity at infinity~[7]. In various models involving the
Chern-Simons term~[4-6], as well as the NLSM model coupled to the Hopf
term~[9], fractional spin had been revealed by essentially computing
the difference between $J^s$ and $J^N$. Since, in the present case
$J^s$ matches exactly with the orbital angular momentum $J^N$, we
conclude that the system (4.1) does not exhibit the existence of
fractional spin in spite of the presence of the Hopf term.
This should not be very surprising considering the fact that the Hopf
term is a total divergence. However,
since this result is purely classical, one cannot rule out the emergence
of fractional spin at the quantum level if Dirac quantization of the
model is carried out.

\subsection{Fractional spin in a radiation-gauge-fixed Hopf-$CP^1$ model}

The result of no fractional spin obtained in the last subsection is
in sharp contradiction to the scenario of NLSM~[9] where the
expression of angular momentum is modified by the presence of an
extra part corresponding to fractional spin emanating from the Hopf
term. The result of fractional spin in [9] is not a typical quantum effect
since it can be obtained in a classical analysis itself which can then
be extended to the quantum level.  Wilczek and Zee~[8] had also
considered the same model, but instead of a Hamiltonian
analysis, they considered a slow adiabatic rotation of the system
by an angle of $2\pi$.  They found the wave function to acquire an
additional phase on this rotation which provided the fractional
spin for the system. It needs to be stressed that the latter way of
obtaining fractional spin is a purely quantum effect.
In this paper we have carried out
our analysis in the Dirac Hamiltonian framework, as in [9]. In fact,
ours is a $CP^1$ version of [9]. We have confined our
analysis to the classical level since transition to the quantum
level by elevating the field dependent DB's (3.9) to quantum
commutators is problematic~[10] because of operator ordering
ambiguities. It is therefore unexpected to disagree with [9].

In what follows we shall show that the Lagrangian
considered in [9] is basically inequivalent to the one (4.1) used by
us. In [8,9] the Lagrangian is
$${\cal L} = {1\over 4} (\p_{\mu} M_a)(\p^{\mu} M_a) - \theta
j^{\mu} A_{\mu} - {\l} (M_a M_a - 1)\eqno(4.48)$$
which is the NLSM coupled to the Hopf term (2.32). At this
stage the identity
$$\int d^2x A_0(x)j_0(x) = - \int d^2x A_i(x) j_i(x)\eqno(4.49)$$
which is valid in the radiation gauge, is used inside the
action~[9] to
reduce the Hopf term in the Lagrangian (4.48) to get
$${\cal L} = {1\over 4} (\p_{\mu} M_a)^2 + 2\theta j_i(x) A_i(x) -
{\l} (M_a M_a - 1)\eqno(4.50)$$

It needs to be noted here that the derivation of the identity
(4.46) requires the inversion of (2.30) to express the gauge field
$A_{\mu}$ in terms of the current $j_{\mu}$ using the radiation
gauge condition. However, the spatial components of (2.30) and (4.24) by
virtue of being relations involving time derivatives ({\it not
constraint equations}) when expressed in terms of the matter fields
$M_a$, are likely to lead to discrepancies in the dynamical
structure of any model if substituted into the original Lagrangian.
Besides, in this case the $j_iA_i$ term is no
longer a total divergence unlike the
pure Hopf term $(\sim j_{\mu}A_{\mu})$ considered in the last section.
Hence, it is improper to hold the original Lagrangian (4.48)
consequent for any result which is obtained after the substitution
of (4.49) into it. In order to clarify this point, let us consider
the gauge {\it variant} $CP^1$ version of the Lagrangian (4.50) given
by
$${\cal L} = \vert D_{\mu} Z \vert^2 + {\theta \over \pi}
\e^{i\nu\l} \p_{\nu} Z^{\dg} \p_{\l} Z Z^{\dg} \p_i Z - \l
(Z^{\dg} Z - 1)\eqno(4.51)$$
where the use of (2.25) and (2.30) has been made. (4.51) can also be
obtained by making use of the identity (4.49) which in terms of the
$Z$ fields looks as
$$\int d^2x Z^{\dg}\dot{Z} \st{\ra}{\nabla}Z^{\dg} \times
\st{\ra}{\nb}Z = \int d^2x Z^{\dg} \st{\ra}{\nb} Z \times
[\st{\ra}{\nb} Z^{\dg} \dot{Z} - \dot{Z}^{\dg}
\st{\ra}{\nb}Z]\eqno(4.52)$$
and is clearly an identity involving time derivatives but {\it not} a
constraint equation.

Note that (4.51)
has certain similarities with (4.1), and can be expressed in the
form of (4.1a) with ${\cal L}_0$ given by (4.1b). ${\cal L}_H$ is
now simplified in the radiation gauge to
$${\cal L}_H = {\theta \over \pi} \e^{i\nu\l} \p_{\nu} Z^{\dg}
\p_{\l} Z Z^{\dg} \p_i Z\eqno(4.53)$$
Nevertheless, (4.53) is first order in time derivative, so that (4.7) is still
valid. The canonically conjugate momenta corresponding to $z_{\a}$
and $z^{*}_{\a}$ are now changed to
$$\pi_{\a} = (D_0 z_{\a})^{*} + {\theta \over \pi} \e^{ij} Z^{\dg}
\p_i Z \p_j z^{*}_{\a}\eqno(4.54)$$
$$\pi^{*}_{\a} = (D_0 z_{\a}) - {\theta \over \pi} \e^{ij} Z^{\dg}
\p_i Z \p_j z_{\a}\eqno(4.55)$$
The rest of the momenta given in (4.4-4.6) remain the same.

It can be checked through the Hamiltonian analysis that
the set of constraints and the symplectic structure given by the
DB's (3.9, 4.16 and
4.17)) remain the same as that of the model (4.1), except for the
Gauss constraint which is now changed to that of the pure $CP^1$
model (4.30). This indicates that the Gauss constraint affects the
$U(1)$ gauge transformation on the $Z$ fields as before (4.19). 
The Lagrangian (4.51) being gauge variant, does not possess this
symmetry. But this is not a serious problem, as (3.9),(4.16) and
(4.17) do not represent the final symplectic structure of the model.
They will undergo further modification when we implement
strongly the Gauss constraint along with the radiation gauge condition.
With that the model actually ceases to be a gauge theory as we
are left with no first class constraint. The exact form of this
final set of DB is however not needed for our purpose, as our
interest is the angular momentum operator.

The symmetric expression for the EM tensor (4.33) for
the model (4.1) undergoes no change in this case since ${\cal L}_H$(4.53)
is still metric independent. The expression for linear momentum is
given by (4.37), and thus the angular momentum is
$$J^s = \int d^2x \e_{ij} x_i [(D_0 Z)^{\dg} (D_j Z) + (D_j
Z)^{\dg} (D_0 Z)]\eqno(4.56)$$
In terms of the phase space variables, and after some
simplification using the strongly valid Gauss constraint (4.30),
$J^s$ reduces to
$$J^s = J^N + {i\theta \over \pi} \e^{pq} \int d^2x \e_{ij}
x_i A_p [(D_j Z)^{\dg} \p_q Z - \p_q Z^{\dg} (D_j Z)]\eqno(4.57)$$
where $J^N$ is the Noether expression of angular momentum. It
can be easily checked that both $J^s$ and $J^N$ generate appropriate
spatial rotation with respect to the DBs (3.9,4.16 and 4.17).
Clearly they will continue to do so with respect to the modified
brackets as well, as the radiation gauge condition preserves
rotational symmetry[5].

The $\theta$- dependent term in (4.57) can be expressed in terms of
the topological charge density $j_0$ (2.30) by using the `strong'
constraint (4.12), as
$$J^s = J^N - 2\theta \int d^2x \e_{ij} x_i A_j
j^0\eqno(4.58)$$
By inverting (2.41), one can write $A_j$ in terms of $j^0$ using the
two-dimensional Green's function $D(x-y)$ satisfying
$$\nabla^2_x D(\stackrel{\rightarrow}{X} -
\stackrel{\rightarrow}{Y}) = \d (\stackrel{\rightarrow}{X} -
\stackrel{\rightarrow}{Y})\eqno(4.59)$$
An explicit form of the Green's function is given by
$$D(\stackrel{\rightarrow}{X}) = {1\over 4\pi} ln \vert
\stackrel{\rightarrow}{X} \vert^2\eqno(4.60)$$
Using this, one can show that (4.58) simplifies to
$$J^s = J^N + \theta {\cal N}^2\eqno(4.61)$$
where ${\cal N}$ is the soliton number and is given by (2.28).
Since $J^N$ is just the orbital angular
momentum, one concludes that the model described by (4.51) exhibits
fractional spin. Let us emphasize, once again, that the 
model (4.51) is basically inequivalent to the model (4.1)
which does not display any fractional spin. Nevertheless, it is
possible to construct a variant of the model (4.48) by making use of
the identity (4.49), which exhibits fractional spin as was shown in
[9]. In this way the possibility of having fractional spin in $2+1$
dimensions as discussed in section 2, is realized.

\vskip 0.5in

\section{Introducing the Chern-Simons term}

In this section we shall investigate the modifications of the symplectic
structure, if any, due to the addition of dynamical terms for the gauge
fields in the form of
a Chern-Simons (CS) term and
compare the analysis with the corresponding
nonrelativistic case~[12]. In [12] the Gauss constraint was
found to be modified in such a manner that the model ceases to be a
$CP^1$ model. In this light, it would be interseting to study the
effect of the CS term on the relativistic model, as well.

The model is
$${\cal L} = (D_{\mu}Z)^{\dg}(D^{\mu}Z) + \theta \e^{\mu\nu\l}
A_{\mu}\p_{\nu}A_{\l} - \l (Z^{\dg} Z - 1)\eqno(5.1)$$
where $D_{\mu} = \p_{\mu} - iA_{\mu}$ is the covariant derivative operator.
The canonically conjugate momenta variables are given by
$$\pi_{\a} = {\d {\cal L} \over \d \dot{z}_{\a}} = (D_0z)^{*}_{\a}
\eqno(5.2)$$
$$\pi^{*}_{\a} = {\d {\cal L} \over \d \dot{z}^{*}_{\a}} = (D_0z)_{\a}
\eqno(5.3)$$
$$\pi^j = {\d {\cal L} \over \d \dot{A}_j} = - \theta \e^{ij}
A_i\eqno(5.4)$$
$$\pi^0 = {\d {\cal L} \over \d \dot{A}_0} = 0\eqno(5.5)$$
$$\pi_{\l} = {\d {\cal L} \over \d \dot{\l}} = 0\eqno(5.6)$$
Note that except for (5.4), all the other relations (5.2-5.6)
appear in case of the pure $CP^1$ model (4.26) also. (5.4) represents a pair of
second class constraints coming from the CS sector which is first
order in time derivative. The relevant DB's which implement (5.4)
`strongly', can either be determined by using the Dirac scheme, or
else can be almost read off by using the symplectic method of
Faddeev and Jackiw~[20], to obtain
$$\{A_i(x), A_j(y)\} = {\e_{ij} \over 2\theta} \d (x-y)\eqno(5.7)$$
The other two constraints (5.5 and 5.6) represent the only two
primary constraints of the model.

The Legendre transformed Hamiltonian is given by
$${\cal H}_c = \pi_{\a}\dot{z}_{\a} + \pi^{*}_{\a}\dot{z}^{*}_{\a}
+ \pi^0 \dot{A}_0 + \pi^j \dot{A}_j + \pi_{\l} \dot{\l} - {\cal
L}$$
$$= \pi^{*}_{\a}\pi_{\a} - iA_0(\pi^{*}_{\a}z^{*}_{\a} -
\pi_{\a}z_{\a}) - 2\theta A_0 \e^{ij} \p_i A_j + \vert D_iZ \vert^2
+ \l (Z^{\dg} Z - 1)\eqno(5.8)$$

Preservation of the primary constraints (5.5 and 5.6) yield the
following secondary constraints
$$G(x) \equiv i(\pi^{*}_{\a}(x)z^{*}_{\a}(x) -
\pi_{\a}(x)z_{\a}(x)) + 2\theta \e^{ij} \p_i A_j(x) \approx
0\eqno(5.9)$$
and
$$C(x) \equiv Z^{\dg}(x)Z(x) - 1 \approx 0\eqno(5.10)$$
The expressions of $\pi_{\a}$ (5.2) and $\pi_{\a}^{*}$ (5.3) are the
same as that of the pure $CP^1$ model. Consequently, the constraint
$$\pi_{\a}^{*}z_{\a}^{*} + \pi_{\a}z_{\a} \approx 0\eqno(5.11)$$
is present here as well. Just as in the $S^3$ model (3.1) the constraints
(5.10) and (5.11) form a second class pair, and the symplectic
structure is given by the DB (3.9)and (5.7).

It can be easily seen that that (5.9) generates the $U(1)$ gauge
transformation. For example
$$\d z_{\a}(x) = \int d^2y f(y) \{z_{\a}(x), G(y)\} = -
if(x)z_{\a}(x)\eqno(5.12a)$$
$$\d A_i(x) = \int d^2y f(y) \{A_i(x), G(y)\} = - \p_i
f(x)\eqno(5.12b)$$
Thus (5.9) can be identified with the Gauss constraint.
Note at this stage that the gauge field $A_i$ is an independent degree
of freedom now. A relation like $A_i = -iZ^{\dg}\p_iZ$ (4.22) is absent
in this case, and therefore, the gauge field has nothing to do with the
monopole connection on $CP^1$. Consequently, the first term
$(D_{\mu}Z)^{\dg}(D^{\mu}Z)$ {\it does not yield} the nonlinear sigma
model. Recall that the equivalence of the $CP^1$ and the NLSM (2.25a)
hinges on two relations, (i) $Z^{\dg}Z = 1$ (5.10), and (ii) $A_{\mu} =
-iZ^{\dg}\p_{\mu}Z$ (2.25b). So just like its nonrelativistic
counterpart where the relation $Z^{\dg}Z =1$ is itself altered~[12],
here too {\it the model ceases to be the} $CP^1$ {\it model}, albiet for
a different reason as (5.10) still holds.

Let us now consider the momentum and the angular momentum generator.
The symmetric expression of the EM tensor is identical to that of the
model with the Hopf term (4.1). Similarly, here also $P_i^s \equiv
\int d^2x T^s_{0i}$ fails to generate appropriate translations ((4.40)
holding true again). Therefore, we modify $T^s_{0i}$ by making the
same choice as (4.43) to get
$${\tilde T}^s_{0i} = \pi_{\a}\p_iz_{\a} +
\pi_{\a}^{*}\p_iz^{*}_{\a} - 2\theta \e^{kl}A_i(\p_kA_l)\eqno(5.13)$$
so that the integrated expression
$${\tilde P}^s_i \equiv \int d^2x {\tilde T}^s_{0i} = \int
d^2x [\pi_{\a}\p_iz_{\a} +
\pi_{\a}^{*}\p_iz^{*}_{\a} - 2\theta \e^{kl}A_i(\p_kA_l)]\eqno(5.14)$$
can now be identified as the momentum generator giving correct
translation. But unlike the case of the Hopf term, this expression of
momentum does not match with the one obtained through the Noether
prescription
$$P_i^N = \int d^2x T^N_{0i} \equiv \int d^2x [\pi_{\a}\p_i z_{\a} +
\pi^{*}_{\a}\p_iz^{*}_{\a} + \pi^j\p_iA_j]$$
$$ = \int d^2x [\pi_{\a}\p_iz_{\a} + \pi^{*}_{\a}\p_iz^{*}_{\a} +
\theta\e_{kl}A_l(\p_iA_k)]\eqno(5.15)$$
by using the `strong' equality (5.4). $P_i^N$ generates appropriate
translations as well, just as in (4.41).

The difference between the Hopf and the CS terms is more striking in
the case of the angular momentum $J$. The expression obtained through
the symmetric EM tensor is given as,
$$J^s=\int d^2x \e^{ij}x_i{\tilde T}^s_{0j}\eqno(5.16)$$
whereas the one using Noether's prescription is given by
$$J^N=\int d^2x [\e^{ij}x_iT^N_{0j}+\pi^i\Sigma^{12}_{ij}A^j]\eqno(5.17a)$$
where,
$$\Sigma^{12}_{ij}=(\d^1_i\d^2_j-\d^1_j\d^2_i)\eqno(5.17b)$$
The presence of an additional $\Sigma$-dependent piece is due to
the presence of an independent dynamical variable $A_i$ which transforms
as a vector under spatial rotation. Note that such a term was absent
in the case of the model involving Hopf term (4.46), as in that case,
the background gauge field $A_i$ ceased to be an independent degree
of freedom because of the strong equality (4.22) relating the gauge
fields $A_i$ to the $Z$ fields. 

The expression (5.17a) which can be simplified as
$$J^N=\int d^2x [\e^{ij}x_iT^N_{0j}+\theta A_jA^j]\eqno(5.18)$$
can be shown to generate appropriate spatial rotation:
$$\left\{Z(x),J^N\right\}=\e_{ij}x_i\p_jZ(x)\eqno(5.19a)$$
$$\left\{A_k(x),J^N\right\}=\e_{ij}x_i\p_jA_k(x)+\e_{ki}A^i\eqno(5.19b)$$
An identical set of equation holds for $J^s$ (5.16) also, indicating
that $J^s$ too generates appropriate rotation. However $J^s$ and
$J^N$ are not identical-they differ by a nontrivial boundary
term $J_b$ given as,
$$J_b=J^s-J^N=\theta \int d^2x \p_i[x_jA^jA^i-x^iA_jA^j]\eqno(5.20)$$
Precisely the same term appears in the context of other models involving
CS term [6,7].Some of its important properties have already been
studied in [6,7], namely that $J_b$ is gauge invariant only under
those gauge transformation, which reduce to identity asymptotically.
Also that the CS gauge field do not fall off to zero asymptotically
fast enough in any of the standard gauges, as can be seen by
looking at the Gauss constraint (5.9). So $J_b$ evaluated in two
different gauges having different asymptotic behaviour is likely
to yield different results. Proceeding as in [7], we can therefore
evaluate this in a rotationally symmetric gauge like radiation
gauge to get 
$$J_b={Q^2 \over {8\pi \theta}}\eqno(5.21)$$
where
$$Q=i\int d^2x (\pi_{\a}(x)z_{\a}(x)-\pi_{\a}^*(x)z_{\a}^*(x))\eqno(5.22)$$
can be interpreted as the total charge of the system. This is
because the integrand corresponds to the zeroth component of the
Euler-Lagrange equation of motion of the gauge field $A_i$.
$$j^{\mu}=\e^{\mu \nu \l}\p_{\nu}A_{\l}\eqno(5.23)$$
where,
$$j_{\mu}=i[(D_{\mu}Z)^{\dg}Z - Z^{\dg}(D_{\mu}Z)]\eqno(5.24)$$
The total charge $Q$ (5.22) can be expressed on the Gauss constraint
surface (5.9) as
$$Q \approx 2\t \int d^2x \e^{ij}\p_i A_j\eqno(5.25)$$
so that this represents the total magnetic flux for the CS gauge field. Since
$A_j$ is no longer equal to $-iZ^{\dg}\p_jZ$, the integrand in (5.25)
cannot be identified with the topological density any more.
A non-zero value of $J_b$ indicates that unlike the Hopf term, the CS term 
imparts an "fractional" spin term right at the classical level.

Here we would like to clarify that the fractional spin in the context
of CS term is conceptually somewhat different from the
corresponding case of Hopf term we have considered earlier. This is because
the Noether's expression for the angular momentum in Hopf case
consists solely of the orbital angular momentum unlike the CS case
(5.17,5.18) which  consists of an additional
spin term apart from the orbital one. The term ``fractional"
in the CS case just indicates the ``anomalous" term we get over and 
above that of the Noether's expression.Also note that the
fractional spin (4.61) in the model involving radiation gauge
fixed Hopf term (4.51) is given in terms of the soliton number
$\cal N$, in contrast to the case here (5.21), where it is given
in terms of the total charge $Q$ (5.22), which in turn is the reflection
of the fact that the CS gauge field $A_i$ has an independent
existence now and has nothing to do with the monopole connection
on $CP^1$, as we have mentioned earlier.

\section{Conclusions}

In this paper we have first investigated some effects on the symplectic
structure due to gauging and
introducing the Hopf term to the $CP^1$
model. To begin with, we
considered the line element on group $SU(2)$ and its coset space
$CP^1 \sim S^2$ and following [13], constructed the $S^3$ model and the $CP^1$
model associated with these respective line elements. The
former enjoys only a global $SU(2)$ symmetry. The $CP^1$ model can be
obtained from this by gauging the $U(1)$ subgroup. We found that
the symplectic structure given by the set of DB's in the $S^3$
model remains unaffected by this. The only difference being the
appearance of a first class (Gauss) constraint generating $U(1)$ gauge
transformation. The $CP^1$ model coupled to the
Hopf term also has identical symplectic structure except that the
structure of the Gauss constraint is modified by an additional
$\theta$-dependent term.

Although a ``background'' gauge field was
introduced in the $CP^1$ model, this however gets related to the
$CP^1$ fields `strongly' so that it effectively becomes a pull back
onto the spacetime of the Dirac magnetic monopole connection on $CP^1
\sim S^2$. This gauge field therefore ceases to be an independent
degree of freedom. This is in contrast to the case where the gauge field
is given dynamics through the Chern-Simons term. The gauge field here is an
independent degree of freedom, and is not related with the monopole
connection of $CP^1$. The symplectic structure of this model also
remains essentially the same, except that the independent CS gauge
fields have an additional Faddeev-Jackiw bracket between themselves.

The fact that the CS gauge field has an independent existence has its
bearing on the Noether expression of angular momentum in the form of
an additional spin term apart from the usual orbital piece. This is in
contrast to the case of the Hopf term where the Noether expression of
angular momentum consists of the orbital part only, arising from the
presence of scalar fields. We have shown that the angular momentum
obtained from the symmetric expression of the EM tensor agrees with
that obtained from the Noether prescription for the $CP^1$ model
coupled to the Hopf term. This indicates that no fractional spin is
imparted at the classical level by the Hopf term, a result that
appears to be in disagreement with that obtained in [9]. One can
attribute this discrepancy to the fact that an identity involving time
derivatives, when expressed in terms of the $CP^1$ variables, is used
to simplify the Hopf term in the radiation gauge. This is
because relations involving time derivatives are not constraint
equations and is liable to alter the dynamical content of the theory
when substituted in the original Lagrangian. To corroborate, we
have also carried out a similar analysis for the radiation gauge
fixed Hopf term coupled to the $CP^1$ model to reveal fractional
spin. This indicates the possibility of having fractional spin,
arising from the nontrivial fundamental group of the
configuration space (2.31), can be realised by (4.50)~[9] or its
$CP^1$ version (4.51), where radiation gauge condition has been
incorporated right at the beginning to construct the model itself.

For the model involving the CS term, one finds that in contrast to the
Hopf case, the two expressions of angular momentum obtained through
the symmetric EM tensor and the Noether prescription differ by a
nontrivial boundary term. This boundary term can be evaluated using
the radiation gauge condition to yield the standard anomalous spin at
the classical level itself. The difference between this result and the
result obtained in the radiation gauge fixed Hopf model is that in the
former case fractional spin is given by the total charge of the
system, whereas in the latter case it is given by the soliton number
$\cal N$.

It would be interesting to quantize the
$CP^1$-Hopf theory in the Dirac scheme, instead of the reduced phase
space scheme, to see if any fractional spin emerges as a pure quantum
effect. Further, it would be also interesting to study if the analysis
can be generalyzed for an arbitrary  compact semi-simple Lie group $G$ and its
coset $G/H$ to see whether the model obtained by gauging
the subgroup $H$ has any effect on the original symplectic structure.

\vskip 0.2in

A.S.M wishes to acknowledge the financial support provided through
a project by the Department of Science and Technology, Government
of India.

\pagebreak

{\bf REFERENCES}

\begin{enumerate}

\item J.F.Schonfeld, Nucl. Phys. B {\bf 185} (1981) 157; S.Deser,
R.Jackiw and S.Templeton, Phys. Rev. Lett. {\bf 48} (1982) 975;
Ann. Phys. (N.Y) {\bf 140} (1982) 372; D.Boyanovsky,
R.Blankenbecker and R.Yabalin, Nucl. Phys. B {\bf 270} (1986) 483;
R.Jackiw, Nucl. Phys. B (Proc. Suppl.) {\bf 18} (1991) 107.
\item E.Fradkin, ``Field Theories in Condensed Matter Systems'',
(Addison Wesley, 1991).
\item S.Forte, Rev. Mod. Phys. {\bf 64} (1992) 193.
\item C.R.Hagen, Ann.Phys.(N.Y.) {\bf 157} (1984) 342; Phys. Rev. {\bf
D 31} (1985) 848; P.K.Panigrahi, S.Roy and W.Scherer, Phys. Rev. Lett.
{\bf 61} (1988) 2827; Phys. Rev. {\bf D38} (1988) 3199; A.Kovner, Phys. Lett.
{\bf B 224} (1989) 299; R.Jackiw, Ann. Phys. (N.Y.) {\bf 201} (1990) 83;
R.Jackiw and S.Y.Pi, Phys. Rev. {\bf D 42} (1990) 3500; G.Semenoff
and P.Sodano, Nucl. Phys. {\bf B 328} (1989) 753; G.Semenoff, Int.
Jour. Mod. Phys. {\bf A7} (1992) 2417.
\item R.Banerjee, Phys. Rev. Lett. {\bf 69} (1992) 17; Phys. Rev. D
{\bf 48} (1993) 2905; R.Banerjee and B.Chakraborty, Phys. Rev. D
{\bf 49} (1994) 5431; B.Chakraborty and A.S.Majumdar, Ann. Phys.
(N.Y) {\bf 250} (1996) 112.
\item R.Banerjee and B.Chakraborty, Ann. Phys. (N.Y) {\bf 247}
(1996) 188; R.Banerjee, Nucl. Phys. B {\bf 419} (1994) 611.
\item B.Chakraborty, Ann. Phys. (N.Y) {\bf 244} (1995) 312.
\item F.Wilczek and A.Zee, Phys. Rev. Lett. {\bf 51} (1983) 2250.
\item M.Bowick, D.Karabali and L.C.R.Wijewardhana, Nucl. Phys. B
{\bf 271} (1986) 417.
\item P.A.M. Dirac, ``Lectures on quantum mechanics'', Belfar
Graduate School of Science (Yeshiva University, New York, 1964);
A.Hanson, T.Regge and C.Teitelboim, ``Constrained Hamiltonian
Ananlysis'' (Academia Nazionale Dei Lincei, Rome, 1976); K. Sundermeyer,
``Constrained Dynamics'', Lecture notes in Physics, Vol 169
(H.Araki et al. Eds.), Springer Verlag, Berlin, 1982; M.Henneaux and
C.Teitelboim, ``Quantization of gauge systems'', Princeton University
Press, Princeton, NJ 1992.
\item Y.S.Wu and A.Zee, Phys. Lett. B {\bf 147} (1984) 325.
\item R.Banerjee and B.Chakraborty, Nucl. Phys. B {\bf 449} (1995)
317.
\item A.P.Balachandran, G.Marmo, B.S.Skagerstam and A.Stern, ``Classical
Topology and Quantum States'' (World Scientific, Singapore, 1991).
\item M.F.Atiyah, ``Geometry of Yang-Mills fields'' (Academia
Nazionale Dei Lincei Senola Normale Superiore, Pisa, 1979).
\item T.Eguchi, P.Gilkey and A.Hanson, Phys. Rep. {\bf 66} (1980)
213; A.Salam and J.Strathdee, Ann. Phys. (N.Y) {\bf 141} (1982) 316.
\item F.J.Hilton,``An introduction to Homotopy theory", Cambridge
University Press, Cambridge, England,1953; N.Steenrod, ``The topology
of fibre bundles", Princeton University Press, 1970; R.Rajaraman,
``Solitons and Instantons",  North-Holland, Amsterdam, 1989.
\item S.Randjbar-Daemi and R.Percacci, Phys. Lett. B {\bf 117}
(1982) 41.
\item B.Chakraborty and A.S.Majumdar, ``On Fractional Spin in the
$CP^1$ model coupled to the Hopf term'', (SNBNCBS preprint, 1996).
\item B.Chakraborty and T.R.Govindarajan, Mod. Phys. Lett. A{\bf 12}
(1997) 619.
\item L.Faddeev and R.Jackiw, Phys. Rev. Lett. {\bf 60} (1988)
1692.

\end{enumerate}

\end{document}